\documentclass[12pt]{article}

\usepackage{graphicx,comment}
\usepackage{amsmath,amssymb,amsfonts,amsthm,comment}

\def\bfm#1{\boldsymbol{#1}}

\let\kappa\varkappa

\let\epsilon\varepsilon

\let\phi\varphi
\let\upn=\textup

\def\etal{{\noexpand\lowercase{\noexpand\normalfont et al.}}}
\def\E{\operatorname{\mathbf E}\nolimits}
\def\P{\operatorname{\mathbf P}\nolimits}

\def\mmid{\mathchoice{\hskip1.5pt|\hskip1.5pt}{\hskip1.5pt|\hskip1.5pt}{\hskip0.5pt|\hskip0.5pt}{\hskip0.3pt|\hskip0.3pt}}
\def\Mmid{\mathchoice{\hskip1.5pt\|\hskip1.5pt}{\hskip1.5pt\|\hskip1.5pt}{\hskip0.5pt\|\hskip0.5pt}{\hskip0.3pt\|\hskip0.3pt}}
\makeatletter
\renewcommand{\pod}[1]{\if@display\mkern10mu\else\mkern6mu\fi(#1)}
\renewcommand{\pmod}[1]{\pod{{\operator@font mod}\mkern6mu#1}}
\makeatother

\renewcommand{\int}{\intop\limits}

\begin{document}

\noindent {\it Problems of Information Transmission},\\
\noindent vol. 58, no. 3, pp. 70--84, 2022.

\begin{center} { M. V. Burnashev} \end{center}

\vskip 0.4cm

\begin{center}
{\large On Minimax Detection of Gaussian Stochastic Sequences with
Imprecisely Known Means and Covariance Matrices} \footnote[1]
{This work was supported  by the Russian Foundation for Basic
Research under Grant 19-01-00364.}
\end{center}



\begin{abstract}
We consider the problem of detecting (testing) Gaussian stochastic sequences
(signals) with imprecisely known means and covariance matrices. The alternative 
is independent identically distributed zero-mean Gaussian random variables with 
unit variances. For a given \emph{false alarm\/} (1st-kind error) probability, 
the quality of minimax detection is given by the best \emph{miss probability\/} 
(2nd-kind error probability) exponent over a growing observation horizon. We 
explore the maximal set of means and covariance matrices (composite hypothesis) 
such that its minimax testing can be replaced with testing a single particular 
pair consisting of a mean and a covariance matrix (simple hypothesis) without 
degrading the detection exponent. We completely describe this maximal set.
\end{abstract}

\section{Introduction and the Main Results}

\subsection{Problem Setting}
One of traditional problems of testing simple hypotheses
$\mathcal{H}_0$ and $\mathcal{H}_1$, concerning \\
Gaussian signal vector
$\boldsymbol{\eta}_n$ in the Gaussian noise background $\boldsymbol{\xi}_n$
(i.e., the problem of signal detection in the noise background), based on 
observations 
$\boldsymbol{y}_n^T=\boldsymbol{y}_n'=(y_1,\ldots,y_n)\in\mathbb{R}^n$ has 
the form
\begin{equation}\label{mod0}
\begin{gathered}
\mathcal H_{0}: {\mathbf y}_{n} = \boldsymbol{\xi}_{n}, \qquad
\boldsymbol{\xi}_{n} \sim
{\mathcal N}(\boldsymbol{0},\mathbf{I}_{n}), \\
\mathcal H_{1}: {\mathbf y}_{n} = \boldsymbol{\eta}_{n}, \qquad
\boldsymbol{\eta}_{n} \sim {\mathcal N}(\mathbf{a}_{n},\mathbf{M}_{n}),
\end{gathered}
\end{equation}
where the sample $\bfm{\xi}_n^T=(\xi_1,\ldots,\xi_n)$ represents ``noise'' and 
consists of independent identically distributed Gaussian random variables 
with zero means and variances $1$, and $\bfm{I}_n$ -- unit covariance matrix. 
Stochastic ``signal''~$\bfm{\eta}_n$ is the Gaussian random variable with 
known mean $\bfm{a}_n$ and known covariance matrix $\bfm{M}_{\!n}$.

However, in practice, we usually do not know precisely the mean $\bfm{a}_n$ and 
the matrix $\bfm{M}_{\!n}$, and then, in reality, the observation model
\eqref{mod0} takes the form
\begin{equation}\label{mod1}
\begin{aligned}
\mathcal{H}_0\colon &\bfm{y}_n=\bfm{\xi}_n,&~ \bfm{\xi}_n &\sim
{\mathcal{N}}(\bfm{0},\bfm{I}_n),\\ \mathcal{H}_1\colon &\bfm{y}_n=\bfm{\eta}_n,&~
\bfm{\eta}_n &\sim {\mathcal{N}}(\bfm{a}_n,\bfm{M}_{\!n}),\quad
\bfm{a}_n\in\mathcal{A}_n,\quad \bfm{M}_{\!n}\in\mathcal{M}_n,
\end{aligned}
\end{equation}
where $\mathcal{A}_n$ -- given set of possible means $\bfm{a}_n$, and
$\mathcal{M}_n$ -- given set of possible covariance matrices
$\bfm{M}_{\!n}$ (probably, depending on $\bfm{a}_n$). We denote for convenience 
$$
\bfm{B}_n=(\bfm{b}_n,\bfm{V}_{\!n}),\quad \bfm{b}_n\in \mathcal{A}_n,\quad
\bfm{V}_{\!n}\in\mathcal{M}_n,\qquad \mathcal{F}_n=\{\bfm{B}_n\}=
(\mathcal{A}_n,\mathcal{M}_n).
$$

Further, for the model \eqref{mod1} we consider the problem of minimax testing
\cite{Wald, Lehmann, Poor} of the simple hypothesis $\mathcal{H}_0$ against 
the composite alternative $\mathcal{H}_1$, based on observations
$\bfm{y}_n^T=\bfm{y}_n'=(y_1,\ldots,y_n)\in\mathbb{R}^n$. If for making 
decision in favor of $\mathcal{H}_0$ a set $\mathcal{D}\in\mathbb{R}^n$ is 
chosen, such that
\begin{equation}\label{testD}
\bfm{y}_n\in\mathcal{D}\ \Rightarrow\ \mathcal{H}_0,\qquad \bfm{y}_n
\not\in\mathcal{D}\ \Rightarrow\ \mathcal{H}_1,
\end{equation}
then the 1st-kind error probability (``\emph{false alarm}'') 
$\alpha(\mathcal{D})$ and the 2nd-kind error \\
probability
(``\emph{miss probability}'') $\beta(\mathcal{D},\mathcal{A}_n,\mathcal{M}_n)$, 
are defined by formulas, respectively,
\begin{equation}\label{defalpha2}
\alpha(\mathcal{D})=\P(\bfm{y}_n \not\in\mathcal{D}\mmid \mathcal{H}_0)
\end{equation}
and
\begin{equation}\label{defbeta2}
\beta(\mathcal{D},\mathcal{A}_n,\mathcal{M}_n)=\sup\limits_{\bfm{a}_n\in\mathcal{A}_n}\,
\sup\limits_{\bfm{M}_{\!n}\in\mathcal{M}_n} \P(\bfm{y}_n\in\mathcal{D}\mmid
\bfm{M}_{\!n},\bfm{a}_n).
\end{equation}

We are interested in the minimal possible 2nd-kind error probability
$\beta(\mathcal{D},\mathcal{A}_n,\mathcal{M}_n)$ (see \eqref{defalpha2} and
\eqref{defbeta2}), provided a given 1st-kind error probability $\alpha$, 
$0<\alpha<1$:
\begin{equation}\label{defbeta3}
\beta(\alpha,\mathcal{A}_n,\mathcal{M}_n)
=\inf\limits_{\mathcal{D}:\:\alpha(\mathcal{D})\le\alpha}
\beta(\mathcal{D},\mathcal{A}_n,\mathcal{M}_n),
\end{equation}
and in the corresponding optimal decision set $\mathcal{D}(\alpha)$ from 
\eqref{testD}.

In the paper, we consider the case when the value $\alpha$ is fixed (or vanishes 
slowly with $n\to\infty$). That case sometimes is called Neyman-Pearson problem 
of minimax testing of hypotheses. In that case the 1st-kind and the 2nd-kind 
errors imply very different losses for the statistician, and he is mainly 
interested in minimization of the 2nd-kind error probability 
$\beta=\P\{\mathcal{H}_0\mmid\mathcal{H}_1\}$.
The case is quite popular in various applications
(see, e.g., \cite{ZhangPoor11} and bibliography therein).

For given mean $\bfm{a}_n$, matrix $\bfm{M}_{\!n}$ and the value $\alpha$ denote 
by $\beta(\alpha,\bfm{a}_n,\bfm{M}_{\!n})$ the minimal possible
2nd-kind error probability (see \eqref{defbeta3}). The corresponding optimal 
decision set $\mathcal{D}(\alpha,\bfm{a}_n,\bfm{M}_{\!n})$ is described by
Neyman -- Pearson lemma \cite{Wald,Lehmann}. Clearly,
\begin{equation}\label{ineq1}
\sup_{\bfm{M}_{\!n}\in\mathcal{M}_n}
\beta(\alpha,\bfm{a}_n,\bfm{M}_{\!n})\le\beta(\alpha,\bfm{a}_n,\mathcal{M}_n).
\end{equation}

For a fixed $\alpha$ and given sets $\mathcal{A}_n,\mathcal{M}_n$, denote also 
by $\beta(\alpha,\mathcal{A}_n,\mathcal{M}_n)$ the minimal possible
2nd-kind error probability (see \eqref{defbeta3}). Then similarly to 
\eqref{ineq1} we have
\begin{equation}\label{ineq1a}
\sup_{\bfm{a}_n\in\mathcal{A}_n} \,\sup_{\bfm{M}_{\!n}\in\mathcal{M}_n}
\beta(\alpha,\bfm{a}_n,\bfm{M}_{\!n})\le\beta(\alpha,\mathcal{A}_n,\mathcal{M}_n).
\end{equation}

In many practical cases the value $\beta(\alpha,\bfm{a}_n,\bfm{M}_{\!n})$ 
decreases exponentially in $n\to\infty$. Therefore, it is natural
(in any case, simpler and more productive) to investigate the corresponding 
exponents $n^{-1}\ln\beta(\alpha,\bfm{a}_n,\bfm{M}_{\!n})$ and
$n^{-1}\ln\beta(\alpha,\mathcal{A}_n,\mathcal{M}_n)$ as $n\to\infty$
(some results on the equality in \eqref{ineq1a} are contained in \cite{Bur17a}).

In the paper, we investigate sets $\mathcal{F}_n=(\mathcal{A}_n,\mathcal{M}_n)$, 
for which in \eqref{ineq1a} the following asymptotic equality holds:
\begin{equation}\label{aseq1a}
\lim_{n\to\infty}\frac{1}{n}\ln \beta(\alpha,\mathcal{A}_n,\mathcal{M}_n)
=\lim_{n\to\infty}\frac{1}{n}\ln\beta(\alpha,\bfm{a}_n,\bfm{M}_{\!n}).
\end{equation}

Motivation for investigation minimax testing of hypotheses (detection
of signals) is described in detail in \cite{Wald, Lehmann, Poor,ZhangPoor11}.
If for given sets of means $\mathcal{A}_n$ and matrices $\mathcal{M}_n$ the 
relation \eqref{aseq1a} holds, then we may replace (without asymptotic losses) 
the entire set $\mathcal{F}_n$ by the particular pair $(\bfm{a}_n,\bfm{M}_{\!n})$. 
Recall that the optimal test for a particular pair $(\bfm{a}_n,\bfm{M}_{\!n})$ is 
described by Neyman -- Pearson lemma and it reduces  to the simple likelihood 
ratio test (LR-test). Otherwise (without relation \eqref{aseq1a}), the optimal 
minimax test is much more complicated Bayes test with respect to 
the \emph{least favorable} prior distribution on the set $\mathcal{F}_n$. 
Therefore, it is natural to investigate when it is possible to replace the 
given set $\mathcal{F}_n$ by a particular pair 
$\bfm{F}_{\!n}=(\bfm{a}_n,\bfm{M}_{\!n})$. But from technical viewpoint 
it is more convenient to consider the equivalent problem: for a given 
pair~$\bfm{F}_{\!n}$ to find the maximal set of pairs 
$\mathcal{F}_n(\bfm{F}_{\!n})$, which can be replaced by the pair 
$\bfm{F}_{\!n}$. This problem is mainly considered in the paper.

{\it Remark 1}.
Models \eqref{mod0} and \eqref{mod1} can be reduced to the
equivalent models with a diagonal matrix $\bfm{M}_{\!n}$. Indeed, since
$\bfm{M}_{\!n}$ -- a covariance matrix (i.e.,~symmetric and positive
definite), there exists an orthogonal matrix $\bfm{T}_{\!n}$ and
a diagonal matrix $\bfm{\Lambda}_n$, such that
$\bfm{M}_{\!n}=\bfm{T}_{\!n}\bfm{\Lambda}_n\bfm{T}_{\!n}'$
(see [\cite{Bellman}, \S\S\,4.7--4.9; \cite{Horn}, Theorem~4.1.5]).
In addition, the diagonal matrix
$\bfm{\Lambda}_n= \bfm{T}_{\!n}'\bfm{M}_{\!n}\bfm{T}_{\!n}$
consists of the eigenvalues $\{\lambda_i\}$ of the matrix $\bfm{M}_{\!n}$. 
Note also that for any orthogonal matrix $\mathbf{T}_{n},$  the vector 
$\bfm{T}_{\!n}'\bfm{\xi}_n$ has the same distribution as that 
of ~$\bfm{\xi}_n$ (for the simple hypothesis $\mathcal{H}_0$ of \eqref{mod1}). 
Therefore, multiplying both sides of \eqref{mod1} by $\bfm{T}_{\!n}'$,
we may reduce the model \eqref{mod1} to the equivalent case with a
diagonal matrix $\bfm{M}_{\!n}$.

{\bf Definition 1}.
For a fixed $\alpha$, and a given sequence of pairs
$\bfm{F}_{\!n}=(\bfm{a}_n,\bfm{M}_{\!n})$ define by
$\mathcal{F}_0(\bfm{F}_{\!n})$ the sequence of the largest sets of pairs, 
such that the equality \eqref{aseq1a} takes the form
\begin{equation}\label{aseq11}
\lim_{n\to\infty}\frac{1}{n}\ln\beta(\mathcal{F}_0(\bfm{F}_{\!n}))
=\lim_{n\to\infty}\frac{1}{n}\ln\beta(\bfm{F}_{\!n}).
\end{equation}
Clearly, $\bfm{F}_{\!n}\in\mathcal{F}_0(\bfm{F}_{\!n})$.

In other words, for a given 1st-kind error probability $\alpha$ the sequence
$\mathcal{F}_0(\bfm{F}_{\!n})$ is the largest set of pairs, which can be 
replaced (without asymptotic losses for $\beta(\mathcal{F}_0(\bfm{F}_{\!n}))$) 
by one pair $\bfm{F}_{\!n}$. Below we describe (Theorem 1) the largest set
$\mathcal{F}_0(\bfm{F}_{\!n})$, satisfying \eqref{aseq11}. It generalizes 
similar result from \cite{Bur20a}, where the case $\bfm{a}_n=\mathbf{0}_n$ was 
considered. It also strengthens similar result from \cite{ZhangPoor11}, where 
for the set $\mathcal{F}_0(\mathbf{0}_n,\bfm{M}_{\!n})$ some lower bounds were 
obtained.

It is convenient first to investigate similar to $\mathcal{F}_0(\bfm{F}_{\!n})$ 
the maximal sets $\mathcal{F}_0^{\rm LR}(\bfm{F}_{\!n})$, which appear if 
LR-detector (see Definition 2) is used. It will be shown that
$\mathcal{F}_0(\bfm{F}_{\!n})=\mathcal{F}_0^{\rm LR}(\bfm{F}_{\!n})$, i.e., 
LR-detector is asymptotically optimal.

In models \eqref{mod0} and \eqref{mod1} denote by $\P_{\bfm{I}_n}$ the 
distribution of the value $\bfm{y}_n= \bfm{\xi}_n$, where
$\bfm{\xi}_n\sim {\mathcal{N}}(\bfm{0},\bfm{I}_n)$. Similarly denote by 
${\mathbf{Q}}_{\mathbf{F}_{n}}$,
$\bfm{F}_{\!n}=(\bfm{a}_n,\bfm{M}_{\!n})$,
the distribution of the value $\bfm{y}_n= \bfm{\eta}_n$, where
$\bfm{\eta}_n \sim {\mathcal{N}}(\bfm{a}_n,\bfm{M}_{\!n})$. 
Denote also by $p^{}_{\bfm{I}_n}(\bfm{y}_n)$ and
$p^{}_{\bfm{F}_{\!n}}(\bfm{y}_n)$, $\bfm{y}_n\in\mathbb{R}^n$, corresponding 
densities of probability distributions. For ($n\times n$)-matrix
$\bfm{M}_n$ denote $|\bfm{M}_n|=\det\bfm{M}_n$. Note that, 
if $|\bfm{M}_{\!n}|\ne 0$, then
\begin{equation}\label{deGaus1a}
\begin{aligned}
\ln p^{}_{\bfm{I}_n}(\bfm{y}_n)&=
-\frac{1}{2}[n\ln(2\pi)+(\bfm{y}_n,\bfm{y}_n)],\quad \bfm{y}_n\in\mathbb{R}^n,\\ 
\ln p^{}_{\bfm{F}_{\!n}}(\bfm{y}_n)&=-\frac{1}{2}
\bigl[n\ln(2\pi)+\ln|\bfm{M}_{\!n}|+(\bfm{y}_n-\bfm{a}_n,
\bfm{M}_{\!n}^{-1}(\bfm{y}_n-\bfm{a}_n))\bigr].
\end{aligned}
\end{equation}
For $|\bfm{M}_{\!n}|\ne 0$ introduce also the logarithm of the
likelihood ratio (see \eqref{deGaus1a})
\begin{equation}\label{deGaus2}
r_{\bfm{F}_{\!n}}(\bfm{y}_n)=
\ln\frac{p^{}_{\bfm{I}_n}}{p^{}_{\bfm{F}_{\!n}}}(\bfm{y})\\[-3pt]\\[-3pt]
=\frac{1}{2}\Bigl[\ln|\bfm{M}_{\!n}|+(\bfm{y}, (\bfm{M}_{\!n}^{-1}-
\bfm{I}_n)\bfm{y}) - 2(\bfm{y},\bfm{M}_{\!n}^{-1}\bfm{a}_n)+(\bfm{a}_n,
\bfm{M}_{\!n}^{-1}\bfm{a}_n)\Bigr].
\end{equation}

Consider first LR-detectors. Introduce the corresponding decision sets
$\mathcal{D}_{\rm LR}(\bfm{F}_{\!n},\alpha)$ in favor of the hypothesis 
$\mathcal{H}_0$ (i.e., in favor of the matrix $\bfm{I}_n$), when simple 
hypotheses $\bfm{I}_n$ and $\bfm{F}_{\!n}$ are tested:
\begin{equation}\label{DLR}
\mathcal{D}_{\rm LR}(\bfm{F}_{\!n},\alpha)=\{\bfm{y}_n\in\mathbb{R}^n:\:
r_{\bfm{F}_{\!n}} (\bfm{y}_n)\ge\gamma\},
\end{equation}
where $\gamma$ is such that, (see \eqref{deGaus2})
\begin{equation}\label{DLR1}
\begin{aligned}[b]
\alpha&=\P_{\bfm{I}_n}\{ \mathcal{D}_{\rm
LR}^{c}(\bfm{F}_{\!n},\alpha)\}=\P_{\bfm{I}_n}\{r_{\bfm{F}_{\!n}}
(\bfm{\xi}_n)\le\gamma\}\\ 
&=\P_{\bfm{I}_n}\Bigl\{ [\ln|\bfm{M}_{\!n}| +(\bfm{\xi}_n,
(\bfm{M}_{\!n}^{-1}- \bfm{I}_n)\bfm{\xi}_n) -
2(\bfm{\xi}_n,\bfm{M}_{\!n}^{-1}\bfm{a}_n)+(\bfm{a}_n,
\bfm{M}_{\!n}^{-1}\bfm{a}_n)]\le 2\gamma\Bigr\}.
\end{aligned}
\end{equation}

{\bf Definition 1}.
For a fixed $\alpha$ and a given sequence of pairs
$\bfm{F}_{\!n}=(\bfm{a}_n,\bfm{M}_n)$ denote by
$\mathcal{F}_0^{\rm LR}(\bfm{F}_{\!n})$ the sequence of the
largest sets of pairs $(\bfm{b}_n,\bfm{V}_{\!n})$, such that
\begin{equation}\label{aseq1aa}
\lim_{n\to\infty}\frac{1}{n}\ln\sup_{(\bfm{b}_n,\bfm{V}_{\!n})\in 
\mathcal{F}_n^{\rm LR}(\bfm{a}_n,\bfm{M}_{\!n})} \beta(\bfm{b}_n,\bfm{V}_{\!n})
=\lim_{n\to\infty}\frac{1}{n}\ln\beta(\bfm{a}_n,\bfm{M}_{\!n}),
\end{equation}
provided the decision sets $\mathcal{D}_{\rm LR}(\alpha,\bfm{a}_n,\bfm{M}_n)$
are used.

Below in Theorem 2 the set $\mathcal{F}_0^{\rm LR}(\bfm{a}_n,\bfm{M}_{\!n})$ for 
the model \eqref{mod1} is described.

We shall also need the following definition \cite{Kullback}.

{\bf Definition 2}.
For probability measures $\mathbf{P}$ and $\mathbf{Q}$ on a measurable space
$(\cal X,\mathcal{B})$ introduce the function (Kullback--Leibler distance 
(or divergence) for measures $\mathbf{P}$ and $\mathbf{Q}$)
\begin{equation}\label{Stein0}
D({\mathbf{P}}\Mmid{\mathbf{Q}})=
\E_{\mathbf{P}} \ln\frac{d\mathbf{P}}{d\mathbf{Q}}(\bfm{x})\ge 0,
\end{equation}
where the expectation is taken over the measure $\mathbf{P}$.

Using formulas \eqref{deGaus1a} and \eqref{Stein0} we have
\begin{equation}\label{deGaus3a}
\begin{aligned}[b]
D({\P_{\bfm{I}_n}}\Mmid{\mathbf{Q}_{\bfm{a}_n,\bfm{M}_{\!n}}})&=
\E_{\bfm{\xi}_n}\ln\frac{p^{}_{\bfm{I}_n}}
{p^{}_{\bfm{a}_n,\bfm{M}_{\!n}}}(\bfm{\xi}_n)\\ 
&=\frac{1}{2}\Bigl[\ln|\bfm{M}_{\!n}|
+(\bfm{a}_n, \bfm{M}_{\!n}^{-1}\bfm{a}_n)+\E_{\bfm{\xi}_n}(\bfm{\xi}_n,
(\bfm{M}_{\!n}^{-1}-\bfm{I}_n)\bfm{\xi}_n)\Bigr]\\
&=\frac{1}{2}\Biggl[\,\sum_{i=1}^n(\ln\lambda_i +\frac{1}{\lambda_i}-1)+
(\bfm{a}_n,\bfm{M}_{\!n}^{-1}\bfm{a}_n)\Biggr],
\end{aligned}
\end{equation}
where $\{\lambda_1,\ldots,\lambda_n\}$ -- the eigenvalues (all positive)
of the covariance matrix $\bfm{M}_{\!n}$, and $\bfm{a}_n=(a_1,\ldots,a_n)$.

\subsection{Assumptions}
In the model \eqref{mod1} denote by
$\lambda_1(\bfm{M}_{\!n}),\ldots,\lambda_n(\bfm{M}_{\!n})$
the eigenvalues (all positive) of the covariance matrix $\bfm{M}_{\!n}$.
We assume that the following assumptions are satisfied:

{\bf I}. For all covariance matrices
$\mathbf{M}_{n} \in {\cal M}_{n}(\mathbf{M}_{n})$
there exists the limit (see \eqref{deGaus3a})
\begin{equation}\label{assump0}
\begin{gathered}
\lim_{n\to\infty}\frac{1}{n}\sum_{i=1}^{n}
\left(\ln\lambda_{i}(\mathbf{M}_{n}) +
\frac{1}{\lambda_{i}(\mathbf{M}_{n})}-1\right)
\end{gathered}
\end{equation}
(note that $\ln z + 1/z-1\geq 0$, $z > 0$).

{\bf II}. For some $\delta > 0$ we have
\begin{equation}\label{assump1}
\begin{gathered}
\lim_{n\to\infty}\frac{1}{n}\sup_{\mathbf{M}_{n} \in {\cal M}_{n}}
\sum_{i=1}^{n}\left|\frac{1}{\lambda_{i}(\mathbf{M}_{n})}-1
\right|^{1+\delta} < \infty.
\end{gathered}
\end{equation}

\subsection{Main results}
We first make an important explanation.

{\it Remark 2}.
There is the following technical problem when describing the maximal sets 
$\mathcal{F}(\bfm{a}_n,\bfm{M}_{\!n})$. The relation \eqref{aseq1a} has the 
asymptotic (as $n\to\infty$) character. Therefore, the maximal sets 
$\mathcal{F}(\bfm{a}_n,\bfm{M}_{\!n})$ can also be described only asymptotically 
(as $n\to\infty$). For that purpose, it is mostly convenient to describe the 
simplest sequence of sets, which gives in the limit
the maximal sets $\mathcal{F}(\bfm{a}_n,\bfm{M}_{\!n})$.

In this paper, for a $n\times n$-matrix $\bfm{A}_n$ we denote 
$|\bfm{A}_n|=\det\bfm{A}_n$. By $(\bfm{x},\bfm{y})$
we denote the inner product of vectors $\bfm{x},\bfm{y}$.
We write $\bfm{A}_n> \mathbf{0}$, if~$\bfm{A}_n$ is positive definite.

Let $\mathcal{C}_n$ -- the set of all $n\times n$-covariance (i.e., symmetric 
and positive definite) matrices in $\mathbf{R}^{n}$. For any
$\bfm{M}_{\!n},\bfm{V}_{\!n}\in\mathcal{C}_n$, and any
$\bfm{a}_n, \bfm{b}_n\in\mathbb{R}^n$ define the function
\begin{equation}\label{deffa}
f_{\bfm{a}_n,\bfm{M}_{\!n}}(\bfm{b}_n,\bfm{V}_{\!n})=\frac{|\bfm{M}_{\!n}|e^{-K}}
{|\bfm{V}_{\!n}\rvert\lvert\bfm{B}_n|},
\end{equation}
where
\begin{equation}\label{Comput2a}
\begin{gathered}
\bfm{B}_n=\bfm{I}_n+\bfm{V}_{\!n}^{-1}-\bfm{M}_{\!n}^{-1},\qquad \bfm{d}=
\bfm{B}_n^{-1}( \bfm{V}_{\!n}^{-1}\bfm{b}_n- \bfm{M}_{\!n}^{-1}\bfm{a}_n),\\
K=(\bfm{b}_n,\bfm{V}_{\!n}^{-1}\bfm{b}_n)- (\bfm{a}_n,
\bfm{M}_{\!n}^{-1}\bfm{a}_n) -
(\bfm{d},\bfm{B}_n\bfm{d}).
\end{gathered}
\end{equation}

For a sequence of pairs $(\bfm{a}_n,\bfm{M}_{\!n})$ introduce the following
sequence of sets of pairs $(\bfm{b}_n,\bfm{V}_{\!n})$:
\begin{equation}\label{Theor1a}
\begin{aligned}[b]
\mathcal{F}_0(\bfm{a}_n,\bfm{M}_{\!n})&=\biggl\{(\bfm{b}_n,\bfm{V}_{\!n}):\:
\E_{\bfm{I}_n}\frac{p^{}_{\bfm{b}_n,\bfm{V}_{\!n}}}
{p^{}_{\bfm{a}_n,\bfm{M}_{\!n}}}(\bfm{x}) \le e^{o(n)}\biggr\}\\
&=\Bigl\{(\bfm{b}_n,\bfm{V}_{\!n}):\: \bfm{I}_n+\bfm{V}_{\!n}^{-1}
-\bfm{M}_{\!n}^{-1}>\mathbf{0},\:
f_{\bfm{a}_n,\bfm{M}_{\!n}}(\bfm{b}_n,\bfm{V}_{\!n})\le e^{o(n)}\Bigr\},
\end{aligned}
\end{equation}
where the function $f_{\bfm{a}_n,\bfm{M}_{\!n}}(\bfm{b}_n,\bfm{V}_{\!n})$ is 
defined in \eqref{deffa}.

The following Theorem is the main result of the paper. It describes the sets \\
$\mathcal{F}(\bfm{a}_n,\bfm{M}_{\!n})$ and
$\mathcal{F}^{\rm LR}(\bfm{a}_n,\bfm{M}_{\!n})$ from \eqref{aseq11} and
\eqref{aseq1aa}, respectively.

{\bf Theorem 1}.
\textit{
If assumptions \/ \eqref{assump0}\upn, \eqref{assump1} hold, then as $n\to\infty$
\begin{equation}\label{Theor1}
\mathcal{F}(\bfm{a}_n,\bfm{M}_{\!n})=\mathcal{F}^{\rm LR}(\bfm{a}_n,\bfm{M}_{\!n})
=\mathcal{F}_0(\bfm{a}_n,\bfm{M}_{\!n}),
\end{equation}
where equalities are understood in the sense of Remark\/ \upn2}.

{\it Remark 3}.
Clearly, $(\bfm{a}_n,\bfm{M}_{\!n})\in\mathcal{F}(\bfm{a}_n,\bfm{M}_{\!n})$. 
Moreover, the sets $\mathcal{F}(\bfm{a}_n,\bfm{M}_{\!n})$ and
$\mathcal{F}^{\rm LR}(\bfm{a}_n,\bfm{M}_{\!n})$ are convex in 
$\bfm{b}_n,\bfm{V}_{\!n}$. Indeed, it is known
[\cite{Bellman}, \S\,8.5,Theorem~4; \cite{Horn}, Theorem~7.6.7],
that the function $f(\bfm{A}_n)= \ln|\bfm{A}_n|$ is strictly concave
on the convex set of positive definite symmetric matrices in $\mathbf{R}^n$. 
Therefore, the set $\mathcal{F}(\bfm{a}_n,\bfm{M}_{\!n})$ is convex, i.e. any 
matrices $\bfm{V}_{\!n}^{(1)}\in\mathcal{F}(\bfm{a}_n,\bfm{M}_{\!n})$ and
$\bfm{V}_{\!n}^{(2)}\in\mathcal{F}(\bfm{a}_n,\bfm{M}_{\!n})$ satisfy condition
$$
a\bfm{V}_{\!n}^{(1)}+(1-a)\bfm{V}_{\!n}^{(2)}
\in\mathcal{F}(\bfm{a}_n,\bfm{M}_{\!n}),\quad \text{for any}\ 0\le a\le 1.
$$

In a sense, $\mathcal{F}(\bfm{a}_n,\bfm{M}_{\!n})$ -- the set 
$\mathcal{F}_0(\bfm{a}_n,\bfm{M}_{\!n})$, enlarged by a ``thin slice''
whose width has the order of $o(n)$. In other words,
$\mathcal{F}_0(\bfm{a}_n,\bfm{M}_{\!n})$ can be considered as a ``core'' of 
the set $\mathcal{F}(\bfm{a}_n,\bfm{M}_{\!n})$.

We present also the following simplifying consequence to Theorem 1. Without
loss of generality, we may assume that the matrix $\bfm{M}_{\!n}$ is diagonal
(see Remark 1) with the eigenvalues $\{\lambda_i\}$ (all positive). We also 
limit ourselves in ~\eqref{Theor1} only to diagonal matrices $\bfm{V}_{\!n}$
with positive eigenvalues $\{\nu_i\}$. The matrix
$\bfm{B}_n= \bfm{I}_n+\bfm{V}_{\!n}^{-1}-\bfm{M}_{\!n}^{-1}$ is diagonal
with the eigenvalues $\{\mu_i\}$:
\begin{equation}\label{deff1}
\mu_i=1+\frac{1}{\nu_i} - \frac{1}{\lambda_i},\quad i=1,\ldots,n.
\end{equation}
Then for $\bfm{a}_n=(a_{1,n},\ldots,a_{n,n})$, 
$\bfm{b}_n=(b_{1,n},\ldots,b_{n,n})$ we have from \eqref{Comput2a}
\begin{equation}\label{deff2}
K=\sum\limits_{i=1}^n\biggl[\frac{b_{i,n}^2}{\nu_i} - 
\frac{a_{i,n}^2}{\lambda_i} -
\frac{1}{\mu_i}\Bigl(\frac{b_{i,n}}{\nu_i} -
\frac{a_{i,n}}{\lambda_i}\Bigr)^2\biggr].
\end{equation}

Introduce the convex set $\mathcal{C}_{{\rm diag},n}$ of diagonal, positive
definite matrices $\bfm{V}_{\!n}$:
$$
\mathcal{C}_{{\rm diag},n}=\{\bfm{V}_{\!n}\in\mathcal{C}_n:\:
\bfm{V}_{\!n}>\mathbf{0}\,\ \text{and}\,\ \bfm{V}_{\!n}\ 
\text{a diagonal matrix}\}.
$$

If $\bfm{M}_{\!n},\bfm{V}_{\!n}\in\mathcal{C}_{{\rm diag},n}$, then the
function $f_{\bfm{a}_n,\bfm{M}_{\!n}}(\bfm{b}_n,\bfm{V}_{\!n})$
from \eqref{deffa} takes the form
\begin{equation}\label{deffVM}
f_{\bfm{a}_n,\bfm{M}_{\!n}}^{(0)}(\bfm{b}_n,\bfm{V}_{\!n})
=e^{-K}\prod_{i=1}^n\Bigl(\frac{\lambda_i}{\nu_i\mu_i}\Bigr),
\end{equation}
where $\{\mu_i\}$ are defined in \eqref{deff1}, and $K$ is defined 
in \eqref{deff2}. It is supposed also, that $\mu_i>0$, $i=1,\ldots,n$.

For a sequence of pairs $(\bfm{a}_n,\bfm{M}_{\!n})$, 
$\bfm{M}_{\!n}\in\mathcal{C}_{{\rm diag},n}$, introduce the following set of
pairs $(\bfm{b}_n,\bfm{V}_{\!n})$,
$\bfm{V}_{\!n}\in\mathcal{C}_{{\rm diag},n}$:
\begin{equation}\label{defV0}
\mathcal{V}(\bfm{a}_n,\bfm{M}_{\!n})=\Bigl\{(\bfm{b}_n,\bfm{V}_{\!n}):\:
1+1/\nu_i-1/\lambda_i>0,\: i=1,\ldots,n,\: \ln
f_{\bfm{a}_n,\bfm{M}_{\!n}}^{(0)}(\bfm{b}_n,\bfm{V}_{\!n})\le o(n)\Bigr\},
\end{equation}
where the function
$f_{\bfm{a}_n,\bfm{M}_{\!n}}^{(0)}(\bfm{b}_n,\bfm{V}_{\!n})$ is defined 
in \eqref{deffVM}.

Then the following ``inner bound'' for $\mathcal{M}(\bfm{a}_n,\bfm{M}_{\!n})$
holds.

{\bf Theorem 2}.
\textit{
If assumptions\/ \eqref{assump0}\upn, \eqref{assump1} hold, then the set
$\mathcal{F}(\bfm{a}_n,\bfm{M}_{\!n})$
contains the set $\mathcal{V}(\bfm{a}_n,\bfm{M}_{\!n})$:
\begin{equation}\label{Cor1}
\mathcal{V}(\bfm{a}_n,\bfm{M}_{\!n}) \subseteq \mathcal{F}
(\bfm{a}_n,\bfm{M}_{\!n}),
\quad \bfm{M}_{\!n}\in \mathcal{C}_{{\rm diag},n},
\end{equation}
where the set\/ $\mathcal{V}(\bfm{a}_n,\bfm{M}_{\!n})$ is 
defined in\/ \eqref{defV0}}.

The set $\mathcal{V}(\bfm{a}_n,\bfm{M}_{\!n})$ is convex in $\bfm{V}_{\!n}$ 
(see Remark 3).

Further, in $\S\,2$ an auxiliary Theorem 3 is given. In $\S\,3$ Theorem~1 is 
proved, and in $\S\,4$ as examples some particular cases of the problem are 
considered.

\section{Auxiliary Theorem }
In models \eqref{mod0}, \eqref{mod1} we first consider the testing of simple 
hypotheses: the pair $(\mathbf{0}_n,{\bfm{I}}_n)$ versus a pair
$(\bfm{a}_n,\bfm{M}_n)$. Denote
$$
D(\mathbf{I}_{n}||\mathbf{a}_{n},\mathbf{M}_{n}) =
D({\mathbf P}_{\mathbf{I}_{n}}||{\mathbf Q}_{\mathbf{a}_{n},\mathbf{M}_{n}}).
$$
Next Theorem is the main auxiliary result of this paper. Its proof follows
the proof of Theorem 3 in \cite{Bur20a}. A more general result is contained 
in \cite{Bur21}.

{\bf Theorem 3}. \textit{
For the minimal possible\/ $\beta(\alpha)$\upn, $0<\alpha<1$\upn, the bounds 
are valid
\begin{equation}\label{Stein11}
\begin{aligned}&
\ln\beta(\alpha)\ge -\frac{D(\mathbf{I}_n\Mmid \bfm{a}_n,\bfm{M}_{\!n})+
h(\alpha)}{1-\alpha},\quad h(\alpha)=-\alpha\ln\alpha - (1-\alpha)\ln(1-\alpha),
\end{aligned}
\end{equation}
and
\begin{equation}\label{Stein12}
\ln\beta(\alpha)\le -D(\bfm{I}_n\Mmid
\bfm{a}_n,\bfm{M}_{\!n})+\mu_0(\alpha,\bfm{a}_n,\bfm{M}_{\!n}),
\end{equation}
where\/ $\mu_0(\alpha,\bfm{a}_n,\bfm{M}_{\!n})$ is defined by the relation}
\begin{equation}\label{defmu0}
\P_{\bfm{I}_n}\biggl\{\ln\frac{p^{}_{\bfm{I}_n}}
{p^{}_{\bfm{a}_n,\bfm{M}_{\!n}}}(\bfm{x})\le D(\bfm{I}_n\Mmid
\bfm{a}_n,\bfm{M}_{\!n})-\mu_0\biggr\}=\alpha.
\end{equation}

Note that both bounds \eqref{Stein11} and \eqref{Stein12} are pure analytical
relations without any limiting operations. The lower bound
\eqref{Stein11} and the upper bound \eqref{Stein12} are close to each other,
if the value $\mu_0(\alpha,\bfm{a}_n,\bfm{M}_{\!n})$ is much smaller than
$D(\bfm{I}_n\Mmid \bfm{a}_n,\bfm{M}_{\!n})$ (which usually has the
order of $n$).

Next result gives an upper bound of the order $n^{1/p}$, $p>1$, for the value
$\mu_0(\alpha,\bfm{a}_n,\bfm{M}_{\!n})$ from \eqref{defmu0}. Its proof
(see Appendix) follows the proof of Lemma 1 in \cite{Bur20a}.

{\bf Lemma \ 1}. \textit{For \/ $\mu_0(\alpha,\bfm{a}_n,\bfm{M}_{\!n})$ \/from 
\eqref{Stein12} \/ the upper bound holds \upn (see \eqref{assump1}\upn)}
\begin{equation}\label{lem1a}
\mu_0(\alpha,\bfm{a}_n,\bfm{M}_{\!n})\le\Biggl(\frac{24}{\alpha}
\sum_{i=1}^n\biggl|\frac{1}{\lambda_i(\bfm{M}_{\!n})}-1\biggr|^p\Biggr)^{1/p}
+3\bigl\|\bfm{M}_{\!n}^{-1}\bfm{a}_n\bigr\|\sqrt{\ln(1/\alpha)}.
\end{equation}

\section{Proof of Theorem 1}
Since $\mathcal{F}_n^{\rm LR}(\bfm{a}_n,\bfm{M}_{\!n}) \subseteq
\mathcal{F}_n(\bfm{a}_n,\bfm{M}_{\!n})$, in order to prove Theorem 1
it is sufficient to get the ``inner bound'' for
$\mathcal{F}_n^{\rm LR}(\bfm{a}_n,\bfm{M}_{\!n})$, and then to get a similar
``outer bound'' for $\mathcal{F}_n(\bfm{a}_n,\bfm{M}_{\!n})$.

\subsection{``Inner bound'' for
$\mathcal{F}_n^{\rm LR}(\bfm{a}_n,\bfm{M}_{\!n})$}
We first estimate from above the value $\beta(\alpha,\bfm{b}_n,\bfm{V}_{\!n})$. 
For that purpose in the model \eqref{mod1} we consider the testing of the simple 
hypothesis $(\mathbf{0}_n,{\bfm{I}}_n)$ against the simple alternative
$(\bfm{a}_n,\bfm{M}_{\!n})$, when $\bfm{a}_n$ is known. We use the optimal
LR-test with the decision region $\mathcal{D}_{\rm
LR}(\bfm{a}_n,\bfm{M}_{\!n},\alpha)=\mathcal{A}_{\mu_0}$ in favor of
$(\mathbf{0}_n,{\bfm{I}}_n)$ (see~\eqref{DLR},~\eqref{DLR1}), where
$\mu_0 =\mu_0(\alpha,\bfm{a}_n,\bfm{M}_{\!n})> 0$ is defined in \eqref{defmu0}. 
Let us consider another pair $(\bfm{b}_n,\bfm{V}_{\!n})$, and evaluate
the 2nd-kind error probability $\beta(\alpha,\bfm{b}_n,\bfm{V}_{\!n})$, 
provided the decision region $\mathcal{A}_{\mu_0}$ is used. Then
\begin{equation}\label{error2V}
\begin{aligned}[b]
\beta(\alpha,\bfm{b}_n,\bfm{V}_{\!n})&=\int\limits_{\mathcal{A}_{\mu_0}}
p^{}_{\bfm{b}_n,\bfm{V}_{\!n}}(\bfm{x})d\bfm{x}
=\int\limits_{\mathcal{A}_{\mu_0}}\frac{p^{}_{\bfm{b}_n,\bfm{V}_{\!n}}}
{p^{}_{\bfm{a}_n,\bfm{M}_n}}(\bfm{x})
\frac{p^{}_{\bfm{a}_n,\bfm{M}_n}}{p^{}_{\bfm{I}_n}}
(\bfm{x})p^{}_{\bfm{I}_n}(\bfm{x})d\bfm{x}\\ &=e^{-D(\bfm{I}_n\Mmid
\bfm{a}_n,\bfm{M}_n)+\mu_1}
\int\limits_{\mathcal{A}_{\mu_0}}\frac{p^{}_{\bfm{b}_n,\bfm{V}_{\!n}}}
{p^{}_{\bfm{a}_n,\bfm{M}_n}} (\bfm{x})p^{}_{\bfm{I}_n}(\bfm{x})d\bfm{x}\\
&\le\beta(\alpha,\bfm{a}_n,\bfm{M}_n)e^{\mu_0}
\E_{\bfm{I}_n}\frac{p^{}_{\bfm{b}_n,\bfm{V}_{\!n}}}
{p^{}_{\bfm{a}_n,\bfm{M}_n}}(\bfm{x}),
\end{aligned}
\end{equation}
where $0\le\mu_1\le\mu_0(\alpha,\bfm{a}_n,\bfm{M}_{\!n})$. Due to the 
assumption \eqref{assump1} and the estimate \eqref{lem1a}, we have
\begin{equation}\label{error2V1a}
\mu_0(\alpha,\bfm{a}_n,\bfm{M}_{\!n})=O(n^{1/(1+\delta)}) =o(n),\quad n\to\infty.
\end{equation}
Therefore, if
\begin{equation}\label{error2V1}
\sup_{(\bfm{b}_n,\bfm{V}_{\!n})\in\mathcal{F}(\bfm{a}_n,\bfm{M}_{\!n})}
\E_{\bfm{I}_n}\frac{p^{}_{\bfm{b}_n,\bfm{V}_{\!n}}} {p^{}_{\bfm{a}_n,\bfm{M}_n}}
(\bfm{x})\le e^{o(n)},\quad n\to\infty,
\end{equation}
then by \eqref{error2V}--\eqref{error2V1} as $n\to\infty$
\begin{equation}\label{error2V2}
\sup_{(\bfm{b}_n,\bfm{V}_{\!n})\in\mathcal{F}(\bfm{a}_n,\bfm{M}_{\!n})}
\ln\beta(\alpha,\bfm{b}_n,\bfm{V}_{\!n})
\le\ln\beta(\alpha,\bfm{a}_n,\bfm{M}_n)+o(n).
\end{equation}

\subsection{``Outer bound'' for $\mathcal{F}_n(\bfm{a}_n,\bfm{M}_{\!n})$}
Now, we get a similar lower bound for $\beta(\alpha,\bfm{b}_n,\bfm{V}_{\!n})$.
Consider first the testing of the simple hypothesis $(\mathbf{0}_n,{\bfm{I}}_n)$ 
against the simple alternative $(\bfm{a}_n,\bfm{M}_n)$. We use the optimal 
LR-test with the decision region 
$\mathcal{D}_{\rm LR}(\bfm{a}_n,\bfm{M}_{\!n},\alpha)=\mathcal{A}_{\mu_0}$ in 
favor of $(\mathbf{0}_n,{\bfm{I}}_n)$ (see~\eqref{DLR},~\eqref{DLR1}). Then, 
denoting $p=p^{}_{\bfm{I}_n}$ and $q=p^{}_{\bfm{a}_n,\bfm{M}_{\!n}}$, we have 
for error probabilities
\begin{equation}\label{Defab11}
\alpha=\P_{\bfm{I}_n}(\mathcal{A}_{\mu_0}),\qquad \beta_{\bfm{a}_n,\bfm{M}_{\!n}}
=\int\limits_{\mathcal{A}_{\mu_0}}q(\bfm{x})\,d\bfm{x}=\beta(\alpha).
\end{equation}

Consider another pair $(\bfm{b}_n,\bfm{V}_{\!n})$. Let 
$\mathcal{D}\in\mathbb{R}^n$ -- a decision region in favor of 
$(\mathbf{0}_n,\bfm{I}_n)$, and
$\beta_{\bfm{b}_n,\bfm{V}_{\!n}}(\mathcal{D})$ and
$\alpha=\alpha(\mathcal{D})$ -- corresponding error probabilities. Then,
denoting $q_1=p^{}_{\bfm{b}_n,\bfm{V}_{\!n}}$, we need to have for the 
2nd-kind error probability $\beta_{\bfm{b}_n,\bfm{V}_{\!n}}(\mathcal{D})$ 
(see \eqref{Defab11})
\begin{equation}\label{Defab1a}
\beta_{\bfm{b}_n,\bfm{V}_{\!n}}(\mathcal{D})
=\int\limits_\mathcal{D}q_1(\bfm{x})\,d\bfm{x}\le\beta(\alpha)e^{o(n)},\qquad
\alpha=\P_{\bfm{I}_n}(\mathcal{D}^{c}).
\end{equation}
For some $\delta$, $0\le\delta\le 1$, consider also the probability density
\begin{equation}\label{Defab2}
q^{}_{\delta}(\bfm{x})=(1-\delta)q(\bfm{x})+\delta q_1(\bfm{x})
\end{equation}
and the corresponding value $\beta_{\delta}$ for it:
\begin{equation}\label{Defab3a}
\beta_{\delta}=\int\limits_\mathcal{D}q^{}_{\delta}(\bfm{x})\,d\bfm{x}
=(1-\delta)\beta(\alpha)+\delta\beta_{\bfm{b}_n,\bfm{V}_{\!n}}.
\end{equation}
We have by \eqref{Defab1a} and \eqref{Defab3a}
\begin{equation}\label{Defab3}
\beta_{\delta}\le\beta(\alpha)\bigl(1-\delta+\delta e^{o(n)}\bigr).
\end{equation}

Note that the probability density $q^{}_{\delta}(\bfm{x})$
corresponds to the Bayes problem statement, when the alternative hypothesis
$\mathcal{H}_1$ with probability $1-\delta$ coincides with
~$(\bfm{a}_n,\bfm{M}_{\!n})$, and with probability $\delta$ -- with 
$(\bfm{b}_n,\bfm{V}_{\!n})$. The value $\beta_{\delta}$ is
the corresponding 2nd-kind error probability.

We lowerbound the value $\beta_{\delta}$. First we have
\begin{equation}\label{Defab4}
\begin{aligned}[b]
\ln\frac{\beta_{\delta}}{1-\alpha}&=\ln\Biggl[\frac{1}{(1-\alpha)}
\int\limits_\mathcal{D}p(\bfm{x})\frac{q^{}_{\delta}}{p}(\bfm{x})\,
d\bfm{x}\Biggr]\ge\frac{1}{(1-\alpha)}
\int\limits_\mathcal{D}p(\bfm{x})\ln\frac{q^{}_{\delta}}{p}(\bfm{x})\,d\bfm{x}\\ 
&=-\frac{D(p(\bfm{x})\Mmid q^{}_{\delta}(\bfm{x}))}{1-\alpha} -
\frac{1}{(1-\alpha)}\int\limits_{\mathcal{D}^{c}}p(\bfm{x})
\ln\frac{q^{}_{\delta}}{p}(\bfm{x})\,d\bfm{x}.
\end{aligned}
\end{equation}
For the last term in the right-hand side of \eqref{Defab4} we have
$$
\int\limits_{\mathcal{D}^{c}}p(\bfm{x})
\ln\frac{q^{}_{\delta}}{p}(\bfm{x})\,d\bfm{x}\le\alpha\ln\Biggl[\frac{1}{\alpha}
\int\limits_{\mathcal{D}^{c}}q^{}_{\delta}(\bfm{x})\,d\bfm{x}\Biggr]
=\alpha\ln\frac{1-\beta_{\delta}}{\alpha}\le\alpha\ln\frac{1}{\alpha}.
$$
Therefore we get
\begin{equation}\label{Stein11a}
\ln\beta_{\delta}\ge -\frac{D(p(\bfm{x})\Mmid
q^{}_{\delta}(\bfm{x}))+h(\alpha)}{1-\alpha}.
\end{equation}
Consider the value $D(p(\bfm{x})\Mmid q^{}_{\delta}(\bfm{x}))$ in the 
right-hand side of \eqref{Stein11a}. Denoting
\begin{equation}\label{Defr}
r(\bfm{x})=\frac{q_1(\bfm{x})}{q(\bfm{x})},
\end{equation}
we have by \eqref{Defab2} and \eqref{Defr}
$$
\frac{q^{}_{\delta}(\bfm{x})}{q(\bfm{x})}=
1-\delta+\delta\frac{q_1(\bfm{x})}{q(\bfm{x})}=1-\delta+\delta r(\bfm{x}).
$$
Therefore
\begin{equation}\label{Defab7}
D(p(\bfm{x})\Mmid q^{}_{\delta}(\bfm{x}))=-\int\limits_{\mathbb{R}^n} p(\bfm{x})
\ln\frac{q^{}_{\delta}}{p}(\bfm{x})\,d\bfm{x}=D(p(\bfm{x})\Mmid
q(\bfm{x}))+g(\delta),
\end{equation}
where
\begin{equation}\label{Defab8}
g(\delta)=- \int\limits_{\mathbb{R}^n} p(\mathbf x)\ln[1-\delta+\delta
r(\bfm{x})]\,d\bfm{x}.
\end{equation}
Therefore, by \eqref{Defab3}, \eqref{Defab7} and \eqref{Defab8} we need to 
have 
\begin{equation}\label{Defab9}
g(\delta)\ge -\ln(1-\delta+\delta e^{o(n)}),\quad \text{for all}\ 0<\delta\le 1.
\end{equation}
Note, that since $\ln \E\xi\ge\E\ln \xi$, then we have from \eqref{Defab8}
$$
g(\delta)\le\ln\int\limits_{\mathbb{R}^n} \frac{p(\bfm{x})}{1-\delta +\delta
r(\bfm{x})}d\bfm{x},\quad \text{for all}\ 0<\delta\le 1.
$$
Therefore, in order to have \eqref{Defab9} fulfilled, we need to have
\begin{equation}\label{Defab92}
\int\limits_{\mathbb{R}^n} \frac{p(\bfm{x})}{1-\delta +\delta
r(\bfm{x})}d\bfm{x}\ge\frac{1}{1-\delta+\delta e^{o(n)}},\quad 0<\delta\le 1.
\end{equation}
Since $\int p(\bfm{x})\,d\bfm{x}=1$, the relation \eqref{Defab92} is equivalent 
to the condition
\begin{equation}\label{Defab92a}
\int\limits_{\mathbb{R}^n} \frac{p(\bfm{x})(r(\bfm{x})-1)}{1-\delta +\delta
r(\bfm{x})}d\bfm{x}\le\frac{e^{o(n)}-1}{1-\delta+\delta e^{o(n)}},\quad 
0<\delta\le 1.
\end{equation}
Note, that
$$
\int\limits_{\mathbb{R}^n} \frac{p(\bfm{x})}{1-\delta +\delta
r(\bfm{x})}d\bfm{x}\le\frac{1}{1-\delta},\quad 0<\delta\le 1.
$$
Then, in order to have \eqref{Defab92a} fulfilled, we need, at least,
\begin{equation}\label{Defab94}
\int\limits_{\mathbb{R}^n} \frac{p(\bfm{x})r(\bfm{x})}{1 +\delta
r(\bfm{x})}d\bfm{x}\le\frac{e^{o(n)}}{(1-\delta)(1-\delta+\delta e^{o(n)})}.
\end{equation}
Setting $\delta \downarrow 0$, we get from \eqref{Defab94} the
necessary condition
\begin{equation}\label{Defab98}
\int\limits_{\mathbb{R}^n}p(\bfm{x})r(\bfm{x})\,d\bfm{x}=\E_{\bfm{I}_n}
\frac{p^{}_{\bfm{b}_n,\bfm{V}_{\!n}}}{p^{}_{\bfm{a}_n,\bfm{M}_n}} (\bfm{x}) \le
e^{o(n)},
\end{equation}
which gives the ``outer bound'' for $\mathcal{F}_n(\bfm{a}_n,\bfm{M}_{\!n})$ 
(see \eqref{Theor1}).

Note that the ``inner bound'' \eqref{error2V1}, \eqref{error2V2} for
$\mathcal{F}_n(\bfm{a}_n,\bfm{M}_{\!n})$ coincides with ~\eqref{Defab98}.
Therefore, in order to finish the proof of Theorem 1 it remains us to express 
analytically the condition \eqref{Defab98} via the matrices
$\bfm{M}_{\!n}, \bfm{V}_{\!n}$ and means $\bfm{a}_n,\bfm{b}_n$. For that 
purpose we use the following result.

{\bf Lemma \ 2}. 
\textit{ If\/ $\bfm{I}_n+\bfm{V}_{\!n}^{-1}-\bfm{M}_{\!n}^{-1}>\mathbf{0}$\upn, 
\/then the formula holds \upn(see \eqref{deffa}--\eqref{Theor1a}\upn)}
\begin{equation}\label{lemma2}
\E_{\bfm{I}_n}\frac{p^{}_{\bfm{b}_n,\bfm{V}_{\!n}}}
{p^{}_{\bfm{a}_n,\bfm{M}_n}}(\bfm{x})=\frac{|\bfm{M}_{\!n}|^{1/2}e^{-K/2}}
{|\bfm{V}_{\!n}|^{1/2}|\bfm{B}_n|^{1/2}}
=f^{1/2}_{\bfm{a}_n,\bfm{M}_{\!n}}(\bfm{b}_n,\bfm{V}_{\!n}),
\end{equation}
\textit{where the function\/ $f_{\bfm{a}_n,\bfm{M}_{\!n}}(\bfm{b}_n,
\bfm{V}_{\!n})$ \/is defined in \eqref{deffa}.}

\textit{If the matrix\/ $\bfm{I}_n+\bfm{V}_{\!n}^{-1}-\bfm{M}_{\!n}^{-1}$\upn 
is not positive definite, then}
\begin{equation}\label{lemma2a}
\E_{\bfm{I}_n}\frac{p^{}_{\bfm{b}_n,\bfm{V}_{\!n}}}
{p^{}_{\bfm{a}_n,\bfm{M}_n}}(\bfm{x})=\infty.
\end{equation}

{\bf Proof}. Denoting
$$
\zeta=\bigl(\bfm{\xi}_n-\bfm{b}_n,\bfm{V}_{\!n}^{-1} (\bfm{\xi}_n-\bfm{b}_n)\bigr)-
\bigl(\bfm{\xi}_n-\bfm{a}_n,\bfm{M}_{\!n}^{-1} (\bfm{\xi}_n-\bfm{a}_n)\bigr),
$$
we get by \eqref{deGaus1a}
\begin{equation}\label{Comput2}
\begin{aligned}[b]
\E_{\bfm{I}_n}\frac{p^{}_{\bfm{b}_n,\bfm{V}_{\!n}}}
{p^{}_{\bfm{a}_n,\bfm{M}_n}}(\bfm{x})
&=\E_{\bfm{\xi}_n}\frac{p^{}_{\bfm{b}_n,\bfm{V}_{\!n}}}
{p^{}_{\bfm{a}_n,\bfm{M}_{\!n}}}(\bfm{\xi}_n)
=\frac{|\bfm{M}_{\!n}|^{1/2}}{|\bfm{V}_{\!n}|^{1/2}} \E_{\bfm{\xi}_n}e^{-\zeta/2}\\
&=\frac{|\bfm{M}_{\!n}|^{1/2}}{|\bfm{V}_{\!n}|^{1/2}
(2\pi)^{n/2}}\int\limits_{\mathbb{R}^n} e^{-\tfrac{1}{2}\left[(\bfm{x},\bfm{x})+
(\bfm{x}-\bfm{b}_n,\bfm{V}_{\!n}^{-1} (\bfm{x}-\bfm{b}_n)) -
(\bfm{x}-\bfm{a}_n,\bfm{M}_{\!n}^{-1} (\bfm{x}-\bfm{a}_n))\right]}\,d\bfm{x}.
\end{aligned}
\end{equation}
Note that (see \eqref{Comput2})
$$
(\bfm{x},\bfm{x})+(\bfm{x}-\bfm{b},\bfm{V}^{-1}
(\bfm{x}-\bfm{b}))-(\bfm{x}-\bfm{a},\bfm{M}^{-1}
(\bfm{x}-\bfm{a}))=(\bfm{x}-\bfm{d},\bfm{B}(\bfm{x}-\bfm{d}))+K,
$$
where (see also \eqref{Comput2a})
$$
\begin{gathered}
\bfm{B}=\bfm{I}+\bfm{V}^{-1}-\bfm{M}^{-1},\qquad \bfm{d}= \bfm{B}^{-1}(
\bfm{V}^{-1}\bfm{b}- \bfm{M}^{-1}\bfm{a}),\\ K=(\bfm{b},\bfm{V}^{-1}\bfm{b})-
(\bfm{a},\bfm{M}^{-1}\bfm{a}) - (\bfm{d},\bfm{B}\bfm{d}).
\end{gathered}
$$
Therefore, we can continue \eqref{Comput2} as follows:
\begin{equation}\label{Comput3b}
\begin{aligned}[b]
\E_{\bfm{I}_n}\frac{p^{}_{\bfm{b}_n,\bfm{V}_{\!n}}}
{p^{}_{\bfm{a}_n,\bfm{M}_n}}(\bfm{x})&=\frac{|\bfm{M}_{\!n}|^{1/2}e^{-K/2}}
{|\bfm{V}_{\!n}|^{1/2}(2\pi)^{n/2}} \int\limits_{\mathbb{R}^n}
e^{-((\bfm{x}-\bfm{d}),\bfm{B}_n(\bfm{x}-\bfm{d}))/2} \,d\bfm{x}\\
&=\frac{|\bfm{M}_{\!n}|^{1/2}e^{-K/2}} {|\bfm{V}_{\!n}|^{1/2}(2\pi)^{n/2}}
\int\limits_{\mathbb{R}^n} e^{-(\bfm{x},\bfm{B}_n\bfm{x})/2}\,d\bfm{x}.
\end{aligned}
\end{equation}
Consider the integral in the right-hand side of \eqref{Comput3b}. If 
$\bfm{B}_n>\mathbf{0}$, then \cite[\S\,6.9, Theorem~3]{Bellman}
\begin{equation}\label{Comput3c}
\int\limits_{\mathbb{R}^n}
e^{-(\bfm{x},\bfm{B}_n\bfm{x})/2}\,d\bfm{x}=\frac{(2\pi)^{n/2}}{|\bfm{B}_n|^{1/2}}.
\end{equation}
Otherwise
\begin{equation}\label{Comput3d}
\int\limits_{\mathbb{R}^n} e^{-(\bfm{x},\bfm{B}_n\bfm{x})/2}\,d\bfm{x}=\infty.
\end{equation}

Assume first $\bfm{B}_n=\bfm{I}_n+\bfm{V}_{\!n}^{-1}-\bfm{M}_{\!n}^{-1}>
\mathbf{0}$, i.e.,
the matrix $\bfm{B}_n$ is positive definite. Then, by \eqref{Comput3b}, 
\eqref{Comput3c} we get
\begin{equation}\label{Stein3b31k}
\E_{\bfm{I}_n}\frac{p^{}_{\bfm{b}_n,\bfm{V}_{\!n}}}
{p^{}_{\bfm{a}_n,\bfm{M}_n}}(\bfm{x})=\frac{|\bfm{M}_{\!n}|^{1/2}e^{-K/2}}
{|\bfm{V}_{\!n}|^{1/2}|\bfm{B}_n|^{1/2}}.
\end{equation}

If the matrix
$\bfm{B}_n=\bfm{I}_n+\bfm{V}_{\!n}^{-1}-\bfm{M}_{\!n}^{-1}$ is not positive 
definite, then by \eqref{Comput3d}
\begin{equation}\label{Stein3b31k1}
\E_{\bfm{I}_n}\frac{p^{}_{\bfm{b}_n,\bfm{V}_{\!n}}}
{p^{}_{\bfm{a}_n,\bfm{M}_n}}(\bfm{x})=\infty,
\end{equation}
and therefore the condition \eqref{Defab98} can not be satisfied. From 
\eqref{Stein3b31k}, \eqref{Stein3b31k1} Lemma 2 follows.\qed

We continue the proof of Theorem 1. Define
$\mathcal{F}(\bfm{a}_n,\bfm{M}_{\!n})$ as the maximal set, satisfying the 
condition
\begin{equation}\label{Stein3b31n}
f_{\bfm{a}_n,\bfm{M}_{\!n}}(\bfm{b}_n,\bfm{V}_{\!n})\le e^{o(n)},\quad n\to\infty.
\end{equation}
That set coincides with the definition \eqref{Theor1a}. Therefore,
from \eqref{error2V1}, \eqref{Defab98}, \eqref{lemma2} and \eqref{Stein3b31n}
Theorem 1 follows.

\section{Examples. Particular cases}

\subsection{Known mean $\bfm{a}_n$ and known covariance matrix $\bfm{M}_{\!n}$}

We first consider the simplest case of known mean $\bfm{a}_n=(a_1,\ldots,a_n)$
and known matrix $\bfm{M}_{\!n}$, and  apply Theorem 3. It will allow us to 
estimate the rate of convergence in Theorem 1. Without loss of generality, we 
may assume in model \eqref{mod1} that the covariance matrix $\bfm{M}_{\!n}$
is diagonal with positive eigenvalues $\lambda_1,\ldots,\lambda_n$
~(see Remark 1). Then (see~\eqref{deGaus3a})
\begin{equation}\label{deGaus5}
\begin{gathered}
D(\mathbf{P}_{\mathbf{I}_{n}}||{\mathbf{Q}}_{\mathbf{a}_{n},\mathbf{M}_{n}}) =
D(\boldsymbol{\xi}_{n}||{\mathbf a}_{n}+\boldsymbol{\eta}_{n}) =
\frac{1}{2}\left[\sum_{i=1}^{n}\left(\ln\lambda_{i} +
\frac{1}{\lambda_{i}}-1 + \frac{a_{i}^{2}}{\lambda_{i}}\right)\right].
\end{gathered}
\end{equation}
By \eqref{Stein11}, \eqref{Stein12} we get for
$D =D(\mathbf{P}_{\mathbf{I}_{n}}||{\mathbf{Q}}_{\mathbf{a}_{n},\mathbf{M}_{n}})$
\begin{equation}\label{Stein17} -\frac{D+1}{1-\alpha}\le\ln\beta(\alpha)\le
-D+\mu_0(\alpha,\bfm{a}_n,\bfm{M}_{\!n}),
\end{equation}
where $\mu_0(\alpha,\bfm{a}_n,\bfm{M}_{\!n})$ is estimated in \eqref{lem1a}.

In order to estimate $\mu_0(\alpha,\bfm{a}_n,\bfm{M}_{\!n})$ simpler than  
\eqref{lem1a}, we assume additionally that the following condition is satisfied:

{\bf III.} There exists $C>0$, such that
\begin{equation}\label{assump3}
\begin{gathered}
{\mathbf E}_{{\mathbf P}}\left[
D(\mathbf{P}_{\mathbf{I}_{n}}||{\mathbf{Q}}_{\mathbf{a}_{n},\mathbf{M}_{n}}) -
\ln\frac{\mathbf{p}_{\mathbf{I}_{n}}}
{\mathbf{p}_{\mathbf{a}_{n},\mathbf{M}_{n}}}
({\mathbf x})\right]^{2} \leq C^{2}
D(\mathbf{P}_{\mathbf{I}_{n}}||{\mathbf{Q}}_{\mathbf{a}_{n},\mathbf{M}_{n}}).
\end{gathered}
\end{equation}

Then by Chebyshev inequality we have
\begin{equation}\label{example1}
\begin{gathered}
\alpha_{\mu} = {\mathbf P}_{\mathbf{I}_{n}}\left\{
D(\mathbf{I}_{n}||\mathbf{a}_{n},\mathbf{M}_{n}) - \ln\frac{p_{\mathbf{I}_{n}}}
{p_{\mathbf{a}_{n},\mathbf{M}_{n}}}({\mathbf x}_{n}) \geq \mu\right\} \leq \\
\leq \mu^{-2}{\mathbf E}_{{\mathbf P}}\left[
D(\mathbf{P}_{\mathbf{I}_{n}}||{\mathbf{Q}}_{\mathbf{a}_{n},\mathbf{M}_{n}}) -
\ln\frac{\mathbf{p}_{\mathbf{I}_{n}}}{\mathbf{p}_{\mathbf{a}_{n},\mathbf{M}_{n}}}
({\mathbf x})\right]^{2} \leq C^{2}\mu^{-2}
D(\mathbf{P}_{\mathbf{I}_{n}}||{\mathbf{Q}}_{\mathbf{a}_{n},\mathbf{M}_{n}}).
\end{gathered}
\end{equation}

In order to have the right-hand side of \eqref{example1} not exceeding $\alpha$, 
it is sufficient to set
$$
\begin{gathered}
\mu = C\sqrt{\frac{D(\mathbf{P}_{\mathbf{I}_{n}}||
{\mathbf{Q}}_{\mathbf{a}_{n},\mathbf{M}_{n}})}{\alpha}},
\end{gathered}
$$
and then \eqref{Stein17} takes the form
$$
-\frac{D+1}{1-\alpha}\le\ln\beta(\alpha)\le -D+C\sqrt{\frac{D}{\alpha}},
$$
which estimates the rate of convergence in \eqref{Stein17}.

Note also that similarly to \eqref{lem1b}, \eqref{lem1c} we can get
\begin{equation}\label{deGaus6}
\begin{gathered}
{\mathbf E}_{{\mathbf P}}\left[
D(\mathbf{P}_{\mathbf{I}_{n}}||{\mathbf{Q}}_{\mathbf{a}_{n},\mathbf{M}_{n}}) -
\ln\frac{\mathbf{p}_{\mathbf{I}_{n}}}
{\mathbf{p}_{\mathbf{a}_{n},\mathbf{M}_{n}}}
({\mathbf x})\right]^{2} =
\frac{1}{2}\sum_{i=1}^{n}\left[\left(1-\frac{1}{\lambda_{i}}\right)^{2} +
2\frac{a_{i}^{2}}{\lambda_{i}^{2}}\right].
\end{gathered}
\end{equation}
Therefore the condition {\bf III} is equivalent to the inequality 
(see \eqref{deGaus5} and \eqref{deGaus6})
$$
\sum_{i=1}^n\biggl[\Bigl(1-\frac{1}{\lambda_i}\Bigr)^2
+2\frac{a_i^2}{\lambda_i^2}\biggr]\le C^2
\Biggl[\sum_{i=1}^n\Bigl(\ln\lambda_i+\frac{1}{\lambda_i}-1
+\frac{a_i^2}{\lambda_i}\Bigr)\Biggr].
$$

{\it Remark 4}.
The assumption \eqref{assump3} is fulfilled, for example, in the 
natural ``regular'' case, when elements $\bfm{a}_{n+1}$, $\bfm{M}_{n+1}$ 
are ``continuations'' of elements~$\bfm{a}_n$,~$\bfm{M}_n$.

\subsection{Unknown mean $\bfm{a}_n$ and known covariance matrix $\bfm{M}_{\!n}$}

Consider the case of model \eqref{mod1},
when we know the covariance matrix $\bfm{M}_{\!n}$, but we do not
know the mean $\bfm{a}_n$. Without loss of generality we may assume
the covariance matrix $\bfm{M}_{\!n}$ diagonal with positive eigenvalues
$\lambda_1,\ldots,\lambda_n$ (see Remark~ 1). Then the function
$f_{\bfm{a}_n,\bfm{M}_{\!n}}(\bfm{b}_n,\bfm{M}_n)$ from \eqref{deffa} takes the 
form
$$
f_{\bfm{a}_n,\bfm{M}_{\!n}}(\bfm{b}_n,\bfm{M}_n)=e^{-K},
$$
where for $\bfm{a}_n=(a_{1,n},\ldots,a_{n,n})$ and 
$\bfm{b}_n=(b_{1,n},\ldots,b_{n,n})$ we have
\begin{equation}\label{example2a}
\begin{aligned}[b]
K&=(\bfm{b}_n,\bfm{M}_{\!n}^{-1}\bfm{b}_n)- (\bfm{a}_n,\bfm{M}_{\!n}^{-1}\bfm{a}_n)-
\bigl(\bfm{M}_{\!n}^{-1}(\bfm{b}_n- \bfm{a}_n), \bfm{M}_{\!n}^{-1}(\bfm{b}_n-
\bfm{a}_n)\bigr)\\ &=\sum\limits_{i=1}^n\biggl[\frac{b_{i,n}^2-a_{i,n}^2}{\lambda_i}-
\frac{(b_{i,n}-a_{i,n})^2}{\lambda_i^2}\biggr].
\end{aligned}
\end{equation}
The corresponding maximal set 
$\mathcal{F}_1(\bfm{a}_n,\bfm{M}_{\!n})=\{\bfm{b}_n\}$ in that case takes the 
form (see \eqref{Theor1a})
\begin{equation}\label{example2b}
\mathcal{F}_1(\bfm{a}_n,\bfm{M}_{\!n})=\{\bfm{b}_n:\: K\ge o(n)\},
\end{equation}
where the function $K=K(\bfm{a}_n,\bfm{M}_{\!n},\bfm{b}_n)$ is defined in 
\eqref{example2a}.

Note that, if $\bfm{M}_{\!n}=\bfm{I}_n$ (i.e., when hypotheses differ only by 
means~$\bfm{a}_n$) formulas \eqref{example2a}, \eqref{example2b} take 
especially simple form:
\begin{equation}\label{example2d}
K=2(\bfm{a}_n,\bfm{b}_n-\bfm{a}_n),\qquad
\mathcal{F}_1(\bfm{a}_n,\bfm{I}_n)=\{\bfm{b}_n:\: (\bfm{a}_n,\bfm{b}_n-\bfm{a}_n) \ge
o(n)\}.
\end{equation}
Those results follow also from papers \cite{Bur79,Bur82} (where that problem
was considered in Hilbert and Banach spaces).

\subsection{Known mean $\bfm{a}_n$ and unknown covariance matrix  $\bfm{M}_{\!n}$}
We limit ourselves to the case $\bfm{a}_n=\mathbf{0}_n$. Then the function 
$f_{\bfm{a}_n,\bfm{M}_{\!n}}(\bfm{b}_n,\bfm{V}_{\!n})$ from
\eqref{deffa} for $\bfm{a}_n=\bfm{b}_n=\mathbf{0}_n$ takes the form
\begin{equation}\label{example3}
f_{\mathbf{0}_n,\bfm{M}_{\!n}}(\mathbf{0}_n,\bfm{V}_{\!n})=
\frac{|\bfm{M}_{\!n}|}{|\bfm{V}_{\!n}|\cdot
\bigl|\bfm{I}_n+\bfm{V}_{\!n}^{-1}-\bfm{M}_{\!n}^{-1}\bigr|}.
\end{equation}
The corresponding maximal set
$\mathcal{F}_1(\mathbf{0}_n,\bfm{M}_{\!n})=\{\bfm{V}_{\!n}\}$
in that case takes the form (see \eqref{Theor1a})
\begin{equation}\label{example3a}
\mathcal{F}_1(\mathbf{0}_n,\bfm{M}_{\!n})=\bigl\{\bfm{V}_{\!n}:\:
f_{\mathbf{0}_n,\bfm{M}_{\!n}}(\mathbf{0}_n,\bfm{V}_{\!n})\le e^{o(n)}\bigr\}.
\end{equation}
Formulas \eqref{example3}, \eqref{example3a} coincide with the corresponding 
results in \cite[Theorem 1]{Bur20a}.

\medskip

\hfill {\large\sl Proof of Lemma 1}
\medskip

Let $\bfm{\xi}_n$ -- a Gaussian random vector with the distribution
$\bfm{\xi}_n \sim {\mathcal{N}}(\bfm{0},\bfm{I}_n)$, and ~$\bfm{A}_n$~-- a 
symmetric $(n \times n)$-matrix with eigenvalues $\{a_i\}$. Consider the 
quadratic form $(\bfm{\xi}_n,\bfm{A}_n\bfm{\xi}_n)$. There
exists the orthogonal matrix $\bfm{T}_{\!n}$, such that
$\bfm{T}_{\!n}'\bfm{A}_n\bfm{T}_{\!n}=\bfm{B}_n$, where
$\bfm{B}_n$ -- the diagonal matrix with diagonal elements~$\{a_i\}$
\cite[\S\,4.7]{Bellman}. Since $\bfm{T}_{\!n}\bfm{\xi}_n \sim
{\mathcal{N}}(\bfm{0},\bfm{I}_n)$, the quadratic forms
$(\bfm{\xi}_n,\bfm{A}_n\bfm{\xi}_n)$ and ~$(\bfm{\xi}_n,\bfm{B}_n\bfm{\xi}_n)$ 
have the same distributions. Therefore, by formula \eqref{deGaus2} we have
\begin{equation}\label{Stein1bb}
\ln\frac{p^{}_{\bfm{I}_n}}{p^{}_{\bfm{a}_n,\bfm{M}_{\!n}}} (\bfm{y}_n)
\stackrel{d}{=} \frac{1}{2}\bigl[\ln|\bfm{M}_{\!n}|+(\bfm{a}_n,
\bfm{M}_{\!n}^{-1}\bfm{a}_n)+\eta_n\bigr],
\end{equation}
where
\begin{equation}\label{Stein1bc}
\eta_n= (\bfm{y}_n, (\bfm{M}_{\!n}^{-1}- \bfm{I})\bfm{y}_n) -
2(\bfm{y}_n,\bfm{M}_{\!n}^{-1}\bfm{a}_n).
\end{equation}

Introduce the value (see \eqref{defmu0})
\begin{equation}\label{lem11}
\alpha_{\mu}=\P_{\bfm{I}_n}\biggl\{\ln\frac{p^{}_{\bfm{I}_n}}
{p^{}_{\bfm{a}_n,\bfm{M}_{\!n}}}(\bfm{x}_n)\le D(\bfm{I}_n\Mmid
\bfm{a}_n,\bfm{M}_{\!n})-\mu\biggr\}.
\end{equation}
Then by \eqref{Stein1bb}, \eqref{Stein1bc} and \eqref{deGaus3a} we have for 
$\alpha_{\mu}$ from \eqref{lem11}
\begin{equation}\label{lem1b}
\begin{aligned}[b]
\alpha_{\mu}&\le \P_{\bfm{\xi}_n}\Biggl\{\Biggl| (\bfm{\xi}_n, (\bfm{M}_{\!n}^{-1}-
\bfm{I})\bfm{\xi}_n) - 2(\bfm{\xi}_n,\bfm{M}_{\!n}^{-1}\bfm{a}_n)-
\sum_{i=1}^n\Bigl(\frac{1}{\lambda_i}-1\Bigr)\Biggr|>2\mu\Biggr\}\\ &=
\P_{\bfm{\xi}_n}\Biggl\{\Biggl|\sum_{i=1}^n\Bigl(\frac{1}{\lambda_i}-1\Bigr)
(\xi_i^2-1)
-2(\bfm{\xi}_n,\bfm{M}_{\!n}^{-1}\bfm{a}_n)\Bigr|>2\mu\Biggr\}\le P_1+P_2,
\end{aligned}
\end{equation}
where
\begin{equation}\label{lem1c}
\begin{aligned}
P_1&=\P_{\bfm{\xi}_n}\Biggl\{\Biggl|
\sum_{i=1}^n\Bigl(\frac{1}{\lambda_i}-1\Bigr)(\xi_i^2-1)\Biggr|>\mu\Biggr\},\\
P_2&=\P_{\bfm{\xi}_n}\bigl\{\bigl|(\bfm{\xi}_n,\bfm{M}_{\!n}^{-1}\bfm{a}_n)\bigl|
>\mu/2\bigr\}.
\end{aligned}
\end{equation}

In order to estimate the value $P_1$ in \eqref{lem1c}, we use the following 
result \cite[Ch.~III.5.15]{Petrov}: let $\zeta_1,\ldots,\zeta_n$ --
independent random variables with $\E\zeta_i=0$, $i=1,\ldots,n$. Then for any 
$1\le p\le 2$
\begin{equation}\label{BarEs1}
\E\Biggl|\sum_{i=1}^n\zeta_i\Biggr|^p\le 2\sum_{i=1}^n\E|\zeta_i|^p.
\end{equation}

Therefore, using for $P_1$ Chebychev inequality and \eqref{BarEs1}, we get
\begin{equation}\label{lem1d}
\begin{aligned}[b]
P_1&\le\mu^{-p}\E\Biggl|\sum_{i=1}^n
\Bigl(\frac{1}{\lambda_i}-1\Bigr)(\xi_i^2-1)\Biggr|^p\le
2\mu^{-p}\sum_{i=1}^n\Bigl|\frac{1}{\lambda_i}-1\Bigr|^p\E |\xi_i^2-1|^p\\ &\le
2\mu^{-p}\sum_{i=1}^n\Bigl|\frac{1}{\lambda_i}-1\Bigr|^p
\bigl(\E|\xi^2-1|^2\bigr)^{p/2}\le 2 \mu^{-p}6^{p/2}
\sum_{i=1}^n\Bigl|\frac{1}{\lambda_i}-1\Bigr|^p\\ &\le
12\mu^{-p}\sum_{i=1}^n\Bigl|\frac{1}{\lambda_i}-1\Bigr|^p.
\end{aligned}
\end{equation}

In order to estimate the value $P_2$ in \eqref{lem1b}, \eqref{lem1c}, note that
$$
(\bfm{\xi}_n,\bfm{M}_{\!n}^{-1}\bfm{a}_n)\sim
\mathcal{N}(0,\|\bfm{M}_{\!n}^{-1}\bfm{a}_n\|),
$$
and then
$$
(\bfm{\xi}_n,\bfm{M}_{\!n}^{-1}\bfm{a}_n) \stackrel{d}{=}
\|\bfm{M}_{\!n}^{-1}\bfm{a}_n\|\xi.
$$
Therefore, using the standard bound
$$
\P(|\xi|\ge z)\le e^{-z^2/2},\quad z\ge 0,
$$
we get ($\xi_i \sim \mathcal{N}(0,1)$)
\begin{equation}\label{lem1f}
P_2=\P_{\bfm{\xi}_n}\bigl\{|
(\bfm{\xi}_n,\bfm{M}_{\!n}^{-1}\bfm{a}_n)|>\mu/2\bigr\}\le
e^{-\mu^2/(8\|\bfm{M}_{\!n}^{-1}\bfm{a}_n\|^2)}.
\end{equation}
In order to satisfy the condition $\alpha_{\mu}\le\alpha$ we set $\mu$, such 
that $\max\{P_1,P_2\}\le\alpha/2$. Then, by \eqref{lem1d} and \eqref{lem1f} it 
is sufficient to set $\mu$, satisfying \eqref{lem1a}.

\section*{FUNDING}

Supported in part by the Russian Foundation for Basic Research, project
no.~19-01-00364.

\end{document}